\documentclass[twocolappendix]{emulateapj}
\usepackage{amsmath}
\usepackage{subfigure}
\usepackage{graphicx}
\usepackage{color}
\usepackage{units}

\newcommand{\V}[1]{ \mathbf #1 }

\newcommand{\Alfven}{Alfv{\'e}n }

\newcommand{\figext}[1]{#1.pdf}

\begin{document}

\title{Crab flares due to turbulent dissipation of the pulsar striped wind}

\author{Jonathan Zrake}
\affil{Kavli Institute for Particle Astrophysics and Cosmology, Stanford
  University, SLAC National Accelerator Laboratory, Menlo Park, CA 94025, USA}

\begin{abstract}

  We interpret $\gamma$-ray flares from the Crab Nebula as the signature of
  turbulence in the pulsar's electromagnetic outflow. Turbulence is triggered
  upstream by dynamical instability of the wind's oscillating magnetic field,
  and accelerates non-thermal particles. On impacting the wind termination
  shock, those particles emit a distinct synchrotron component $F_{\nu,\rm
    flare}$, which is constantly modulated by intermittency of the upstream
  plasma flow. Flares are observed when the high-energy cutoff of $F_{\nu,\rm
    flare}$ emerges above the fast-declining nebular emission around 0.1 - 1
  GeV. Simulations carried out in the force-free electrodynamics approximation
  predict the striped wind to become fully turbulent well ahead of the wind
  termination shock, provided its terminal Lorentz factor is $\lesssim 10^4$.

\end{abstract}

\maketitle

\section{Introduction} \label{sec:introduction}

Discovery of $\gamma$-ray flares from the Crab Nebula is among the foremost
contributions to recent high-energy astrophysics. While it has now been five
years since their announcement in 2011 \citep{Tavani2011, Abdo2011}, no longer
wavelength counterparts have been established \citep{Bietenholz2014, Kouzu2013,
  Bietenholz2014, Rudy2015, Madsen2015}, and so the active region within the
nebula remains unlocalized. The flares also present a formidable theoretical
challenge because their power and duration cannot be accounted for within
conventional theories of charged particle acceleration
\citep[e.g.][]{Blandford2015a, Blandford2015}. Evidently, they are telling us
something new --- either about the nebula's anatomy, or the physics of strongly
magnetized plasma, or perhaps both.

Here we suggest Crab flares are the signature of an intermittent pulsar
wind. Seen in this way, they provide empirical support for a turbulence
resolution to the so-called $\sigma$-problem, referring to uncertainty over the
mechanism by which the pulsar's electromagnetic spin-down luminosity is diverted
into particles \citep[e.g.][]{Kennel1984a, Emmering1987}. Pulsar wind plasma
thus inherits spatial and temporal intermittencies that are characteristic of
relativistic turbulence \citep{Zrake2011, Zrake2012, Radice2013}, and flares can
be produced when exceptionally large coherent structures transit the wind
termination shock.

Our central assertion is that free energy associated with the pulsar's
alternating current (``striped wind'') destabilizes quickly, and is dissipated
in the ensuing turbulent cascade. This view is supported by recent advances in
the stability and reconnection processes of strongly magnetized plasma,
\citep{Uzdensky2010, Cerutti2014, Sironi2014, Guo2015, East2015, Zrake2015}, but
differs with the conventional view that the striped wind is erased by steady
magnetic reconnection through current layers that remain near equilibrium
\citep{Coroniti1990, Lyubarsky2001, Kirk2003}.

Essential features of a model follow from this assertion. First, we envision
that synchrotron radiation is produced by supra-thermal electrons emerging from
the wind zone into the compressed downstream magnetic field, so that photons may
exceed the classical $\sim \unit[100]{MeV}$ limit throughout the duration of
electron cooling times. Second, we envision those electrons to be energized by
turbulent dissipation in the flow well upstream of the wind termination
shock. Finally, we attribute observed $\gamma$-ray variabilities to spatial
intermittency of the upstream electron population; flares are seen when a gust
of wind particles (henceforth a ``blob'') sweeps across the wind termination
shock along the line of sight. We will show that a model having these basic
features can account for the flare duration and energetics without invoking
explosive conversion of magnetic energy into radiation.

In Section \ref{sec:free-energy} we summarize previous work on states of
magnetized plasma resembling the striped wind. In Section
\ref{sec:linear-instabilities} we estimate that current layers in the wind
succumb quickly to linear instabilities. Then in Section \ref{sec:simulations} we
simulate their subsequent non-linear evolution in the force-free electrodynamics
approximation, finding that stripes are fully engulfed by turbulence after about
two comoving dynamical times. In Section \ref{sec:causality} we discuss freely
decaying turbulence in a relativistically expanding background flow, and
determine the maximum scale of coherent structures emerging from the wind
region. In Section \ref{sec:model} we use those results to develop a simplistic
model for the Crab Nebula flares. The model requires a wind Lorentz factor
$\Gamma \approx 2300$, a post-shock magnetic field strength of $\unit[655]{mG}$,
and can be falsified by observation of flares about 5 times more powerful than
the April 2011 event. In Section \ref{sec:discussion} we discuss limitations of
the model and follow-up work that may validate certain assumptions.

\section{Turbulence transition in the striped wind} \label{sec:turbulence}

\begin{figure*}
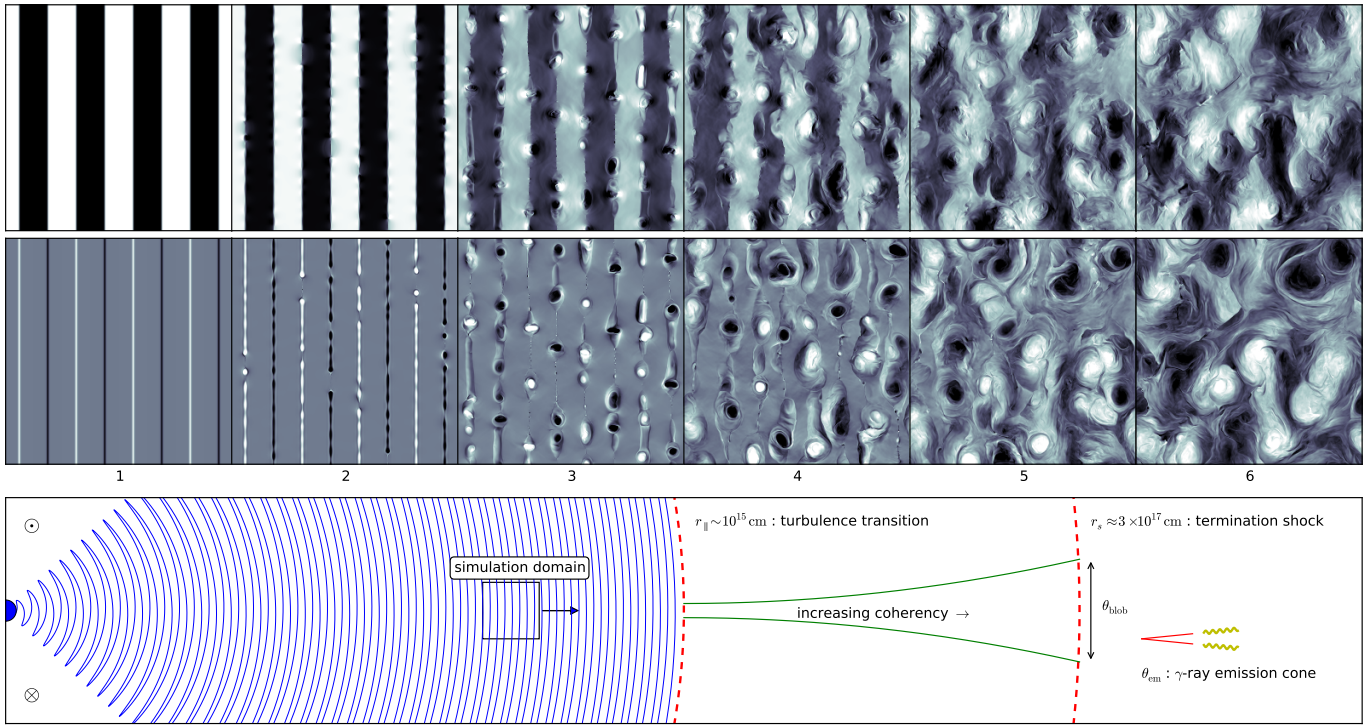

  \centering
  \includegraphics[width=7.1in]{\figext{fig_images_B2}}\vspace{-4pt}
  \includegraphics[width=7.1in]{\figext{fig_images_B3}}\vspace{-4pt}
  \includegraphics[width=7.1in]{\figext{fig_diagram}}
  \caption{Transition to turbulence in the pulsar striped wind. Simulations were
    carried out for a local patch of wind in the plasma rest frame, using the
    force-free electrodynamics approximation. Shown are relief plot renderings
    made from two-dimensional subsets of three-dimensional data, with the
    toroidal (in-page) magnetic field component in the top panel, and the
    poloidal (out-of-page) component in the middle panel. The left-most column
    shows the initial condition given by Equation \ref{eqn:current-layer-ic}.
    The second column is shown just after saturation of a linear instability,
    and subsequent columns illustrate erasure of the stripes by transition to
    fully developed turbulence. Since the magnetic field weakens from left to
    right by turbulent dissipation, each image uses a color bar that is scaled
    to the instantaneous range of magnetic field values. The bottom panel (not
    to scale) shows a schematic diagram of the pulsar (blue circle), rippled
    current sheet (blue line), comoving simulation domain (square), and relative
    locations of the turbulence transition region $r_\parallel$ and termination
    shock radius $r_s$.}
  \label{fig:images}
\end{figure*}

\subsection{Free energy supply} \label{sec:free-energy}

Turbulence in the pulsar wind feeds on the oscillating magnetic field, or
``stripes''. The stripes arise by rippling of the equatorial current sheet as
the pulsar's magnetic dipole vector circles its spin axis, and are generic to
plasma winds sourced by oblique rotators \citep{Parker1958, Coroniti1990,
  Bogovalov1999}. As seen in the local rest frame of wind particles, they are an
abundant source of magnetostatic free energy \footnote{A system's magnetostatic
  free energy is what it can dissipate while respecting the frozen-in assumption
  ($\V E \cdot \V B = 0$) in all but infinitesimal volumes. Such evolution
  conserves the system's ideal topological invariants in the sense of
  \cite{Taylor1974}, while permitting conversion of magnetic energy into bulk
  motions and eventually heat.}. In earlier work, we found that such ``excited''
states of magnetized plasma were dynamically unstable \citep{East2015}, and
promptly discharged their free energy through a turbulent cascade
\citep{Zrake2015}. The force-free equilibria ($\V J \times \V B = 0$) we had
considered were the short-wavelength Taylor states \citep{Childress1970,
  Dombre1986} which have uniform torsion $\alpha \equiv \V J \cdot \V B /
B^2$. Although stripes are different in that $\alpha$ is non-uniform (electrical
current concentrates into thin layers around which the toroidal field switches
sign), we will see in Section \ref{sec:simulations} that they similarly tend
toward dynamical instability.

\subsection{Linear instabilities} \label{sec:linear-instabilities}

Turbulence can be induced by saturation of any small amplitude
instability. Here, we estimate the growth rate of linear tearing modes affecting
current layers \citep[e.g.][]{Biskamp1986} between domains of opposite magnetic
polarity. The upshot is that tearing modes saturate fast, so our conclusions
would not change if another instability grows faster. Current layers in high
Lundquist number plasma tend to form plasmoids \citep{Bhattacharjee2009,
  Huang2010, Uzdensky2010}, or become strongly corrugated \citep{Inoue2012}. Our
estimate here thus predicts how long current layers remain near equilibrium
before such non-linear effects set in.

The fastest growing tearing mode has a wavelength comparable to the comoving
layer thickness $\tilde{a}$, and a growth rate $c \tilde{\lambda}_D^{-3/2}
\tilde{a}^{1/2}$ when the plasma is hot \citep{Zelenyi1979}, where
$\tilde{\lambda}_D$ is the relativistic Debye length. Current layers collapse
down to microscopic width $\tilde{a} = \tilde{\lambda}_D$ \citep{Michel1994}, so
$\tilde{\omega}_{\rm tear} = c / \tilde{a}$. As measured in the pulsar frame the
current layer thickness near the light cylinder was found by \cite{Uzdensky2014}
to be $a_L \approx \unit[30]{cm}$. Plasma near the base of the wind is launched
outward with a Lorentz factor $\gamma_L \approx 200$ in the case of the Crab
pulsar \citep{Lyubarsky2001}, so the comoving current layer thickness $\tilde{a}
= \gamma_L a$ is tens of meters. The tearing rate is a factor $\gamma_L$ slower
in the pulsar frame, and it slows further as the current layer inflates $\propto
r$ (charge density $n_e \propto r^{-2}$ while $\lambda_D \propto n_e^{-1/2}$),
\begin{equation}
  \omega_{\rm tear} = \gamma_L^{-2} \frac{c}{a_L} \left( \frac{r}{r_L}
  \right)^{-1}. 
\end{equation}
Near the light cylinder, $\omega_{\rm tear} \sim \unit[10^4]{s^{-1}}$. By
comparison, the Crab pulsar angular frequency $\Omega \approx 190$. In principle
the tearing mode could be suppressed by inflation of the layer. However, the
ratio of $\omega_{\rm tear}$ to the layer expansion rate $\omega_{\rm exp} = c /
r$ is given by $\gamma_L^{-2} r_L / a_L \sim 100$ independent of distance, so
expansion may be safely neglected. The same also implies that tearing modes
complete $\sim 100$ exponential foldings in the vicinity of the wind's base. We
conclude that linear instability only characterizes the striped wind at
extremely small radii $r \ll r_s$, and that evolution throughout the vast
majority of its flight to the termination shock takes place in the non-linear
regime.

\subsection{Simulations} \label{sec:simulations}

In order to illustrate non-linear evolution of the striped wind, we have carried
out time-dependent numerical simulations in the plasma rest frame. We assume
that the wind's $\sigma$ parameter is sufficiently high that the force-free
electrodynamics (FFE) approximation is appropriate. FFE corresponds to the
limiting case of relativistic magnetohydrodynamics (MHD) where electromagnetic
contributions to the stress-energy tensor dominate those of matter. Use of
continuum rather than fully kinetic formalism is justified by the same argument
given in Section \ref{sec:linear-instabilities}, namely that each wavelength
contains millions to billions of plasma skin depths. FFE is formulated as
Maxwell's equations, together with the Ohm's law
\begin{equation} \label{eqn:ffe-ohm}
  \V J = \frac{\V B}{B^2}\left( \V B \cdot \nabla \times \V B - \V E \cdot
  \nabla \times \V E \right) + \frac{\V E \times \V B}{B^2} \rho,
\end{equation}
that follows from imposing perfect conductivity $\V E \cdot \V B = 0$ and the
force-free condition $\rho \V E + \V J \times \V B = 0$
\citep[e.g.][]{McKinney2006a}. We use a numerical scheme that is fourth order
accurate in space and time. Details are given in \cite{Zrake2015}.

\begin{figure}
  \centering
  \includegraphics[width=3.55in]{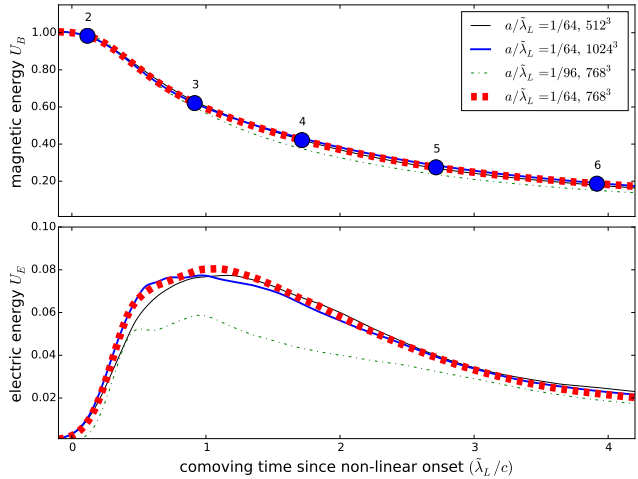}
  \caption{Time series of the integrated magnetic (top) and electric (bottom)
    energy in the simulation depicted in Figure \ref{fig:images}. Time is
    measured since the onset of non-linear evolution, in units of the comoving
    light-travel time between stripes $\tilde{\lambda}_L / c$. Different curves
    show the same model at different grid resolution. The numbered blue circles
    correspond to the columns of Figure \ref{fig:images}.}
  \label{fig:tseries}
\end{figure}

Current layers in a striped wind collapse to an equilibrium supported by gas
pressure as in the solution of \cite{Harris1962}. Since FFE neglects gas
pressure, we instead construct our initial data as a sequence of force-free
rotational current layers, that are in equilibrium by virtue of uniform magnetic
pressure ($B^2 = \rm{const}$) and vanishing tension density ($\V B \cdot \nabla
\V B = 0$). All such force-free equilibrium fields may be parameterized by the
angle $\phi(x)$ of the magnetic field vector in the transverse plane, as a
function of longitudinal position $x$ (where $x$ is measured as a fraction of
the domain length $N_{\rm layer} \tilde{\lambda}_L / 2$); $\V B = B_0 (\cos
\phi(x), \sin \phi(x), 0)$. In order to model current layers having width $a$,
we choose $\phi(x)$ to increase over the domain in a series of $N_{\rm layer}$
steps,
\begin{equation} \label{eqn:current-layer-ic}
  \phi(x) = \pi \sum_{n=1}^{2 N_{\rm layer}} \Theta_a{ \left( x - \frac{2 n -
      1}{2 N_{\rm layer}} \right) } ,
\end{equation}
where $\Theta_a(x) = \frac{1}{2}\left( 1 + \tanh x/a \right)$ is a ``smoothed''
Heaviside step function. Even though our current layers are force-free rather
than supported by gas pressure, they still admit a small-amplitude instability
not unlike the tearing mode. A full description of that instability is outside
the present scope; for our purposes it suffices to say that its growth rate was
seen to increase as $a$ decreases and that saturation is reached after a few
dynamical times $a/c$. Once non-linear effects set in, the limited range of
length scales becomes tolerable due to universality of the turbulent cascade
with respect to small-scale closure, which we will give evidence for
momentarily.

Our simulations are run on the triple periodic domain of size $N_{\rm layer}
\tilde{\lambda}_L / 2$, where $N_{\rm layer} = 8$ and the grid resolution was
$512^3$, $768^3$, or $1024^3$. The current layer width $a$ was varied between
$1/64$ and $1/96$ of the stripe wavelength $\tilde{\lambda}_L$. We neglect
transverse stretching, as well as longitudinal acceleration that may arise from
heating of the plasma \citep[e.g.][]{Kirk2003}. Simulations are initiated in the
force-free equilibrium state given by Equation \ref{eqn:current-layer-ic}, where
the electric field is zero apart from a low amplitude ($10^{-6}$) white-noise
perturbation introduced to break translational symmetry along the layers.

Figure \ref{fig:images} contains renderings of two-dimensional slices of the
toroidal (in-page, top panel) and poloidal (out-of-page, middle panel) magnetic
field component. The left-most column shows the initial condition as described
in Equation \ref{eqn:current-layer-ic}. The second column depicts the system
when the linear instability first gives way to large-amplitude effects. By the
third column, an ensemble of plasmoids (flux tubes oriented out of the page) has
emerged. An important feature of these structures is that they would be
long-lived were translational symmetry along their axis to be imposed, as was
shown in \cite{Zrake2015}. Without imposing that symmetry, the flux tubes are
unstable to kink and sausage modes. They also tend to coalescence with other
flux tubes whose current flows in the same direction. This latter effect
exemplifies the mechanism of non-helical inverse energy transfer studied in
\cite{Zrake2014}. Subsequent columns portray the transition to fully developed
turbulence. The whole sequence takes place over roughly 4 comoving light-travel
times $\tilde{\lambda}_L / c$ of a stripe, which is also the elapsed time
experienced by plasma elements in transit from pulsar to termination shock when
their Lorentz factor is $\sim 10^4$. In Figure \ref{fig:tseries} we plot the
time series of volume-averaged magnetic and electric energy for different grid
resolutions. Relative consistency among different parameters indicates that the
rate of energy dissipation by turbulence is insensitive to the current layer
width and grid resolution, supporting the view that evolution in the non-linear
regime is universal with respect to unresolved physics.

\subsection{Causality} \label{sec:causality}

Transition to turbulence occurs when plasmoids grow to the scale of the stripe
separation $\tilde{\lambda}_L$. Our results from Section \ref{sec:simulations}
indicate this occurs (fourth column of Figure \ref{fig:images}) after a comoving
fast magnetosonic time $\tilde{\lambda}_L / \tilde{v}_f$, that is when causal
contact between stripes is first established. For a wind that moves with
constant velocity $v_w$, this occurs at the radius \footnote{The expression for
  $r_\parallel$ in Equation \ref{eqn:r-parallel} reflects the distance at which
  two fast magnetosonic waves first meet, one propagating downstream starting at
  $r=0$, and another in the upstream direction from $r=\lambda_L$.}
\begin{equation} \label{eqn:r-parallel}
  r_\parallel = \frac{1}{2} \lambda_L \frac{v_w}{\tilde{v}_f} \left( \frac{1 -
    v_w^2 \tilde{v}_f^2 / c^4}{1 - v_w^2/c^2} \right).
\end{equation}
As shown in Figure \ref{fig:tseries}, about half the magnetic energy has been
dissipated by the time plasmoids reach the stripe scale, so prior evolution
occurs in the regime where $\tilde{v}_f$ is only somewhat smaller than $c$ and
the flow is super-fast magnetosonic, $\tilde{\gamma}_f \ll \Gamma$ (Lorentz
factors corresponding to $\tilde{v}_f$ and $v_w$). In that limit, Equation
\ref{eqn:r-parallel} gives us $r_\parallel \approx \Gamma^2 \lambda_L$. During
this phase, the flow is likely to accelerate outward due to loss of inward
tension provided by the toroidal magnetic field, and also by establishing a
turbulent pressure gradient. A detailed analysis along the lines of
\cite{Lyubarsky2001}, together with a turbulence closure of the MHD equations
would thus be necessary to determine the acceleration profile. Here, we adopt
the conservative approximation that $v_w$ appearing in Equation
\ref{eqn:r-parallel} corresponds to the terminal wind Lorentz factor. Transition
to turbulence would thus be avoided by causality if $\Gamma$ were to be $\gtrsim
10^4$. Faster winds would reach the deceleration point before a signal travels
between adjacent current layers, so corrugations could not grow large enough to
effect mixing between them. In Section \ref{sec:model} we will constrain
$\Gamma$ empirically within our interpretation of the Crab flares.

The third column of Figure \ref{fig:images} shows our prediction for the stripe
morphology when the wind passes through $r_\parallel$, that is after a comoving
fast magnetosonic time has elapsed since the onset of non-linear
evolution. Beyond a few times $r_\parallel$, the wind evolves as freely decaying
turbulence of magnetized relativistic plasma. The fully developed turbulent
cascade could in principle be slowed by transverse expansion of the flow, so we
need to repeat the exercise of Section \ref{sec:linear-instabilities}, now
comparing the eddy turn-over frequency $\omega_e = v_A / \lambda_e$ with the
expansion rate $\omega_{\rm exp} = c / r$. Here, we have invoked that turbulent
motions decay alongside the \Alfven speed $v_A$ \citep{Zrake2014}.  As seen in
the plasma rest frame (Figure \ref{fig:images}), eddies are initially isotropic
and roughly the size of the comoving stripe separation $\tilde{\lambda}_L$. They
are thus seen in the pulsar frame to be stretched by a factor $\Gamma$ in the
transverse direction. Since we wish to compare transverse stretching rate with
the eddy frequency, we assign eddies the transverse scale, $\lambda_e = \Gamma
\lambda_L$. Keeping in mind that turbulent cells increase in size $\propto r$
due to the expansion, as well as by inverse energy transfer \citep{Zrake2014,
  Brandenburg2015, Olesen2015}, the eddy size is parameterized by $\lambda_e =
\Gamma \lambda_L (r/r_\parallel)^{1+\delta}$ where $\delta \ge 0$ determines the
rate at which turbulence increases its coherence scale due to inverse energy
transfer alone. The ratio of eddy turnover to expansion is thus
\begin{equation}
  \frac{\omega_e}{\omega_{\rm exp}} = \Gamma \left(\frac{\sigma}{1 +
    \sigma}\right)^{1/2} \left(\frac{r}{r_\parallel}\right)^{-\delta},
\end{equation}
where we have related $\sigma$ to the \Alfven speed and assumed that
$\tilde{v}_f$ and $v_w$ are both effectively $c$ out to $r_\parallel$. Thus, at
least out to a few $r_\parallel$, expansion can be safely neglected, and our
simulation results (which do not account for that expansion) remain applicable.

What if turbulence evolves to increase its coherency at the fastest rate allowed
by causality? Suppose that two fluid elements on radial trajectories with
Lorentz factor $\Gamma$ are initially at radius $r_0$ and separated by an angle
$\theta$. A pulse of light emitted by one is first received by the other when
they have moved to a radius $r_1 \approx r_0 \Gamma^2 \theta^2$,
provided\footnote{The exact answer is $r_1/r_0 = f + \sqrt{f^2 - 1}$ where $f =
  \Gamma^2 \left( 1 - \cos \theta \right) + \cos \theta$.} $\Gamma \gg
\theta^{-1}$. By assuming that turbulence commences at $r_0 = r_\parallel$ and
setting $r_1 = r_s$, we determine the maximum angular coherence scale at the
termination shock to be
\begin{equation} \label{eqn:causality-scale}
  \theta_{e, \rm max} \sim \left( \frac{r_s}{\lambda_L} \right)^{1/2} \Gamma^{-2} .
\end{equation}

\section{A model for the Crab Nebula $\gamma$-ray flares} \label{sec:model}

Here we develop a simplistic model for the Crab Nebula's observed $\gamma$-ray
variability \citep{Tavani2011, Abdo2011, Balbo2011, Buehler2012, Striani2013,
  Rudy2015}. We adopt a tentative interpretation of the April 2011 event in
which the rise time $\tau_{\rm rise} \approx \unit[10]{hr}$ coincides with the
emergence of a giant coherent structure (or ``blob'') from the upstream wind
into the post-shock flow, while the decline $\tau_{\rm dec} \approx
\unit[2]{days}$ is associated with the cooling time of the blob's highest energy
particles.

We assume the shock's kinetic structure to be mediated by energy-bearing
particles having Lorentz factor $\Gamma$, while those of highest energy
$\gamma_{\rm max} \gg \Gamma$ are only weakly deflected through an angle
$\theta_{\rm def}$ before cooling in the post-shock magnetic field $B_s$. The
angular scale $\theta_{\rm em}$ from which photons are received is the larger of
$\gamma_{\rm max}^{-1}$ and $\theta_{\rm def}$, and is reliably the latter. We
assume that only a small patch of the blob surface is visible, $\theta_{\rm
  blob} > \theta_{\rm em}$, so that the flare's inferred isotropic luminosity is
independent of $\theta_{\rm em}$. A further assumption is that turbulence
develops similar longitudinal and transverse coherency in the pulsar rest
frame. Though harder to justify (eddies typically emerge from the wind pancaked
by Lorentz contraction), this condition only needs to be fulfilled
intermittently both in time and solid angle. By setting $\theta_{\rm blob} =
\theta_{e, \rm max}$ (Equation \ref{eqn:causality-scale}), and $r_s \theta_{\rm
  blob} = c \, \tau_{\rm rise} \approx \unit[10^{15}]{cm}$ where $r_s \approx
\unit[3 \times 10^{17}]{cm}$, the bulk Lorentz factor is determined to be
\begin{equation}
  \Gamma \approx \left( \frac{r_s}{c \, \tau_{\rm rise}} \right)^{1/2} \left(
  \frac{r_s}{\lambda_L} \right)^{1/4} \approx 2.3 \times 10^3.
\end{equation}
Next, we assume that particles emerge from the wind with a power-law
distribution $f_{\rm flare}(\gamma) \propto \gamma^{-p}$ between $\Gamma$ and
$\gamma_{\rm max}$. We adopt the spectral index $p = 3/2$ found in kinetic
simulations of relativistic magnetic reconnection \citep{Sironi2014}. The total
number of emitting particles is given by
\begin{equation}
  \varepsilon_e \dot N \tau_{\rm rise} = \int_{\Gamma}^{\gamma_{\rm max}}{f_{\rm
      flare}(\gamma) d \gamma}
\end{equation}
where $\varepsilon_e$ is the fraction of $\dot N$ that became non-thermal within
the blob. Equating the flare duration $\tau_{\rm dec} \approx \unit[2]{days}$
with the cooling time of electrons moving with $\gamma_{\rm max}$ in the
post-shock magnetic field, we find $B_s \approx \unit[1.7]{mG} \, \gamma_{\rm
  max,9}^{-1/2}$. The remaining free parameters are $\gamma_{\rm max}$ and the
rate $\varepsilon_e \dot N$ of non-thermal particles emerging from the wind with
an angle $\theta_{\rm em}$ around the line of sight. They are determined by
fitting Fermi-LAT data at the peak of the April 2011 flare to the synchrotron
photon spectrum $F_{\nu, \rm flare}$ of the flaring component,
\begin{equation}
  F_{\nu, \rm{flare}} = \int_{\Gamma}^{\gamma_{\rm max}} P_\nu f_{\rm
    flare}(\gamma) d \gamma
\end{equation}
where $P_\nu$ is the specific synchrotron power per unit energy
\citep{Rybicki1979}. We then evolve $f_{\rm flare}(\gamma)$ by synchrotron
cooling in the post-shock magnetic field to determine the spectrum at later
times.

\begin{figure}
  \centering
  \includegraphics[width=3.55in]{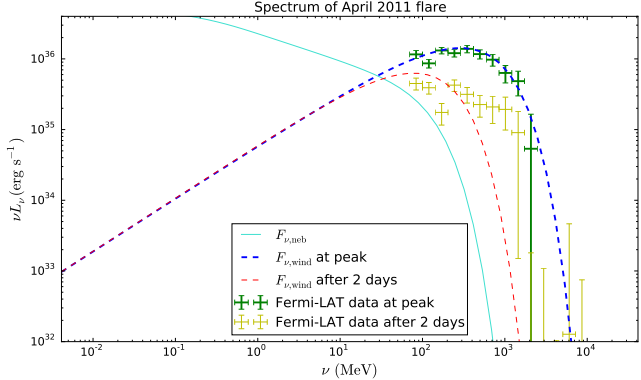}
  \caption{The quiescent Crab Nebula synchrotron spectrum, shown together with
    the flaring component shortly after the peak of the April 2011 event, and
    then another 2 days later.}
  \label{fig:spectrum}
\end{figure}

\begin{figure}
  \centering
  \includegraphics[width=3.55in]{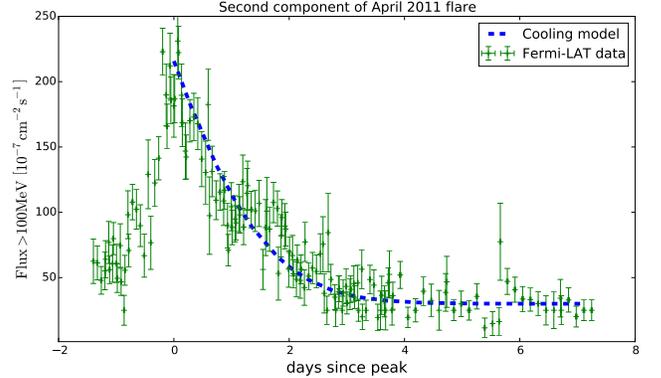}
  \caption{Decline of the second component of the April 2011 flare. Shown is the
    Fermi-LAT photon flux above 100 MeV between MJD 55666 and 55674, together
    with the model with $B_s = \unit[655]{\mu G}$ and $\gamma_{\rm max} = 7
    \times 10^{9}$.}
  \label{fig:lightcurve}
\end{figure}

Figure \ref{fig:spectrum} shows $F_{\nu, \rm flare}$ fitted to the peak of the
April 2011 flare, using Fermi-LAT data around MJD 55667 \citep{Buehler2012},
while Figure \ref{fig:lightcurve} shows the Fermi-LAT photon flux light curve
above 100 MeV together with post-shock cooling of $f_{\rm flare}$. The best-fit
magnetic field value $B_s = \unit[655]{\mu G}$ is somewhat higher than the mean
nebular field of $\unit[124]{\mu G}$ \citep{Meyer2010}, but that is not
unexpected in such close proximity to the shock. The corresponding maximum
Lorentz factor is $\gamma_{\rm max} = 7 \times 10^9$. The best-fit injection
rate of non-thermal particles is $\varepsilon_e \dot N = \unit[2.2 \times
  10^{37}]{s^{-1}}$, which corresponds to an average non-thermal power supply of
$\unit[5.3 \times 10^{37}]{erg/s}$ throughout $\tau_{\rm rise}$ and within an
angle $\theta_{\rm blob}$ of the line of sight. The former is $22\%$ of the
pulsar's conventionally adopted production rate $\dot N \sim
\unit[10^{38}]{s^{-1}}$ \citep{Rees1974, Coroniti1990}\footnote{Particle
  production rates $\gtrsim \unit[10^{40}]{s^{-1}}$ for the Crab pulsar are
  inferred assuming the nebula electron population derives directly from the
  pulsar \citep{Bucciantini2011}.}, while the latter is $11\%$ of the isotropic
pulsar spin-down power $L = \unit[4.6 \times 10^{38}]{erg/s}$
\citep{Komissarov2012}. A bulk Lorentz factor $\Gamma \approx 2300$ is much
smaller than the \cite{Kennel1984a} value of $\sim 10^6$, but comparable to that
of \cite{Tanaka2010} who estimate $\Gamma \approx 7 \times 10^3$. Kinetic
studies of relativistic reconnection in strongly magnetized plasma agree on
values of $\varepsilon_e \sim 1$ \citep{Guo2015, Sironi2014}.

If the post-shock magnetic field were to be coherent over $c \, \tau_{\rm cool}
\approx \unit[5 \times 10^{15}]{cm}$, then particles moving with $\gamma_{\rm
  max}$ would be deflected through an angle $\omega_g \tau_{\rm cool} \approx
10^2 \, \theta_{\rm blob}$. Thus downstream trajectories must be in the Bohm
limit to satisfy our assumption that $\theta_{\rm em} < \theta_{\rm
  blob}$. However, magnetic field correlations would need to vanish at scales
$\lambda_B \gtrsim \unit[10^6]{cm}$ for radiation to be in the jitter regime,
$\gamma_{\rm max}^{-1} < \omega_g \lambda_B / c$ \citep{Kelner2013}.

\section{Discussion} \label{sec:discussion}

We have argued that the Crab Nebula's $\gamma$-ray variability can be induced by
turbulent intermittency of the pulsar wind. Our analysis predicts that linear
instability of the current layers gives way to turbulence well upstream of the
wind termination shock, provided the bulk Lorentz factor is $\lesssim 10^4$. We
went on to estimate the largest spatial coherency attainable by freely decaying
turbulence in the radially expanding flow, and found that the $\sim
\unit[10]{hr}$ rise time of the April 2011 event can be accounted for if $\Gamma
\approx 2300$. We then fit a simplistic model to the Fermi-LAT spectrum of the
flare, and determined roughly 1/10 of the pulsar's spin-down luminosity must be
channeled into non-thermal particles to explain it.

\subsection{Limitations}

Our numerical treatment here was limited to a local comoving patch of wind
plasma. A more realistic analysis will need to account for expansion in the
radial background flow, which could modify the nature of linear
instabilities. We have also ignored all kinetic processes, which may complicate
arguments given in Section \ref{sec:linear-instabilities} that linear
instabilities act quickly. Still, we note that kinetic particle-in-cell
simulations \citep{Sironi2011, Hoshino2012} also point to large-scale disruption
of striped magnetic fields not unlike the behavior in force-free electrodynamics
simulations seen here.

We have also approximated the pulsar wind as having constant Lorentz factor as a
function of radius, even though one dimensional MHD solutions involve
acceleration due to both ideal \citep[e.g.][]{Michel1973} and dissipative
\citep[e.g.][]{Lyubarsky2001} processes. Such acceleration would follow the
turbulence transition (due to loss of the inward tension force provided by the
toroidal field), leading to suppression of causal contact and thus limiting the
growth of coherent structures, which are crucial to our interpretation of the
Crab Nebula flares. However, we have also ignored the prospect that current
layers disrupt at close distance to the pulsar, when the Lorentz factor is still
$\lesssim 10^2$. That would allow turbulence more time over which to develop
spatial coherency. Our analysis of the tearing instability in Section
\ref{sec:linear-instabilities} implies that appearance of plasmoids in the
vicinity of the light cylinder is certainly possible. Actually, such behavior is
an essential feature of the pulsed GeV emission scenario for the Crab pulsar
sketched in \cite{Uzdensky2014}. Large-amplitude kinking of the current layer
near the light cylinder in axially symmetric aligned rotator solutions has also
been reported by \cite{Cerutti2015}.  Such early disruption would imply rapid
acceleration very close to the wind's base. Interestingly, abrupt acceleration
of the wind at around $30 r_L$ has been inferred from the Crab's pulsed emission
in the very high energy $\gamma$-rays by \cite{Aharonian2012}.

\subsection{Radiative efficiency of the wind zone}

Very little radiation will be produced in the wind zone, even if it becomes
turbulent. The reason is seen to be that comoving synchrotron power $\propto
\tilde{B}^2 \tilde{\gamma}$ is suppressed by a factor of $\Gamma^2$, while the
travel duration is shortened by another factor of $\Gamma$. Specifically, a
single particle traveling from radius $r_0$ to the termination shock with
Lorentz factor $\gamma$ will radiate a fraction $\varepsilon_{\rm rad}$ of its
kinetic energy given by
\begin{equation}
  \int_{r_0}^{r_s} \omega_{\rm sync} \, dt = \frac{2}{3}
  \frac{r_e^2r_LB_L^2}{m_e c^2} \Gamma^{-3} \gamma \left( \frac{r_L}{r_0} -
  \frac{r_L}{r_s} \right)
\end{equation}
where $\omega_{\rm sync} = P_{\rm sync} / \gamma m_e c^2$, $r_e$ is the
classical electron radius, and the magnetic field $B_L (r/r_L)^{-1}$ is the
upper limit corresponding to no dissipation. Assuming that $B_L \approx
\unit[10^6]{G}$ and that radiation begins around the turbulence transition $r_0
= r_\parallel$, we find that
\begin{equation}
  \varepsilon_{\rm rad} < 2.1 \times 10^{-6} \left( \frac{\gamma_{\rm max}}{7
    \times 10^9} \right) \left( \frac{\Gamma}{2300} \right)^{-5} .
\end{equation}

\subsection{Comparison with other models}

Our model for Crab flares has several features in common with the ideas of other
authors. For example, the ``magnetic untwisting'' scenario of
\cite{Sturrock2012} also envisions energy release to originate around
$\unit[10^{13}-10^{15}]{cm}$, well within the wind zone. While our analysis
indicates that significant dissipation of magnetic energy is possible there,
poor radiative efficiency implies that any flaring emission could not be
produced \emph{in situ}. \cite{Teraki2013} have also developed a model in which
flares occur when blobs emerge from the wind zone, and examined the possibility
that radiation occurs in the jitter regime. Our analysis provides a physical
basis for the production of such blobs, but we find the condition for jitter
radiation, namely that post-shock magnetic coherency peaks at $\sim
\unit[10^6]{cm}$ (Section \ref{sec:model}), to be quite strict, as the
relaxation time for such small-scale turbulence is extremely
short. \cite{Bykov2012} proposed the flares correspond to rarest excursions in
magnetic field intensity of plasma moving through the termination shock, which
elicit radiative enhancements from an otherwise stationary electron
population. That scenario also looks favorable if the pulsar wind contains
turbulence. More conclusive evidence regarding possible TeV counterparts may
help determine whether radiative intermittency originates in the particle
distribution or magnetic field \citep{Bednarek2011}. Future analysis aiming to
measure the spatial statistics of magnetic versus non-thermal energy density in
decaying plasma turbulence could also be fruitful.

Another avenue that looks encouraging if the inner wind contains turbulence is
one in which flares occur when the termination shock normal moves through the
line of sight due to its own intrinsic fluctuations \citep{Camus2009}. Such a
scenario may treat values of $\varepsilon_e \dot{N}$ and $\gamma_{\rm max}$ we
inferred in Section \ref{sec:model} as reflecting \emph{continuous} wind
parameters, and that intermittency is introduced by fluctuations in the angle of
the post-shock flow. We opted to explore the ``blob'' model for this work
because of the appearance that Crab flare light curves are not time-symmetric,
it because it allowed us to exploit the connection between the wind Lorentz
factor and turbulence coherence scale discussed in Section
\ref{sec:causality}. If further data indicates equal rise and decline times, or
if the rather low Lorentz factor implied by our model can be independently ruled
out, then the fluctuating termination shock approach seems to be a good
choice. At least, it benefits from high efficiency of particle acceleration
expected if turbulence operates throughout the far upstream flow.

Flares have also been suggested to occur by spontaneous magnetic reconnection
events occurring somewhere in the nebula \citep{Uzdensky2011, Clausen-Brown2012,
  Cerutti2014, Cerutti2014a}. While relativistic magnetic reconnection is now
understood to energize non-thermal particles with high efficiency, the April
2011 event requires \emph{all} the magnetic energy (if released isotropically)
in a region $\unit[\sim 10^{16}]{cm}$ with $\sim \unit[]{mG}$ intensity to be
promptly converted into \emph{fast-cooling} electrons. Thus, special
reconnection geometries (``mini-jets'') in which the particle exhaust is
confined to a very narrow opening angle must be invoked. \cite{Lyubarsky2012}
proposed a model involving quasi-cyclic instabilities near the base of the polar
jet, which seems to be supported by full-scale simulations of the Crab Nebula
\citep{Porth2013}. However, if the flares occur by discharge of accumulated
magnetic energy, their power could sometimes exceed that of the pulsar, which
has not been seen yet (meanwhile, our model would be ruled out by such
observations). Arguments \citep[e.g.][]{Clausen-Brown2012} in favor of
``accumulation-discharge'' models appeal to the extreme stability of the pulsar
as the nebula's power supply. Be that as it may, wind plasma is an unsteady
transmission line for delivery of the pulsar's AC power to the nebula.

\acknowledgments The author gratefully acknowledges Jon Arons for inspiring
discussions, and thoughtful contributions from Mikhail Belyaev, Krzysztof
Nalewajko, Yajie Yuan, Roger Blandford, Lorenzo Sironi, Jonathan Granot, Andrew
MacFadyen, and Tom Abel. Simulations were run on the Comet cluster at the San
Diego Supercomputer Center (SDSC) through XSEDE grant AST150038, as well as
Pleiades of the NASA High-End Computing (HEC) Program through the NASA Advanced
Supercomputing (NAS) Division at Ames Research Center.

\bibliographystyle{apj}

\begin{thebibliography}{}
\expandafter\ifx\csname natexlab\endcsname\relax\def\natexlab#1{#1}\fi

\bibitem[{Abdo {et~al.}(2011)Abdo, Ackermann, Ajello, Allafort, Baldini,
  Ballet, Barbiellini, Bastieri, Bechtol, Bellazzini, Berenji, Blandford,
  Bloom, Bonamente, Borgland, Bouvier, Brandt, Bregeon, Brez, Brigida, Bruel,
  Buehler, Buson, Caliandro, Cameron, Cannon, Caraveo, Casandjian, {\c{C}}elik,
  Charles, Chekhtman, Cheung, Chiang, Ciprini, Claus, Cohen-Tanugi, Costamante,
  Cutini, D'Ammando, Dermer, de~Angelis, de~Luca, de~Palma, Digel, {do Couto e
  Silva}, Drell, Drlica-Wagner, Dubois, Dumora, Favuzzi, Fegan, Ferrara, Focke,
  Fortin, Frailis, Fukazawa, Funk, Fusco, Gargano, Gasparrini, Gehrels,
  Germani, Giglietto, Giordano, Giroletti, Glanzman, Godfrey, Grenier, Grondin,
  Grove, Guiriec, Hadasch, Hanabata, Harding, Hayashi, Hayashida, Hays, Horan,
  Itoh, J{\'{o}}hannesson, Johnson, Johnson, Khangulyan, Kamae, Katagiri,
  Kataoka, Kerr, Kn{\"{o}}dlseder, Kuss, Lande, Latronico, Lee,
  Lemoine-Goumard, Longo, Loparco, Lubrano, Madejski, Makeev, Marelli,
  Mazziotta, McEnery, Michelson, Mitthumsiri, Mizuno, Moiseev, Monte, Monzani,
  Morselli, Moskalenko, Murgia, Nakamori, Naumann-Godo, Nolan, Norris, Nuss,
  Ohsugi, Okumura, Omodei, Ormes, Ozaki, Paneque, Parent, Pelassa, Pepe,
  Pesce-Rollins, Pierbattista, Piron, Porter, Rain{\`{o}}, Rando, Ray, Razzano,
  Reimer, Reimer, Reposeur, Ritz, Romani, Sadrozinski, Sanchez, {Saz
  Parkinson}, Scargle, Schalk, Sgr{\`{o}}, Siskind, Smith, Spandre, Spinelli,
  Strickman, Suson, Takahashi, Takahashi, Tanaka, Thayer, Thompson, Tibaldo,
  Torres, Tosti, Tramacere, Troja, Uchiyama, Vandenbroucke, Vasileiou,
  Vianello, Vitale, Wang, Wood, Yang, \& Ziegler}]{Abdo2011}
Abdo, A.~A., Ackermann, M., Ajello, M., {et~al.} 2011, Science (New York,
  N.Y.), 331, 739

\bibitem[{Aharonian {et~al.}(2012)Aharonian, Bogovalov, \&
  Khangulyan}]{Aharonian2012}
Aharonian, F.~A., Bogovalov, S.~V., \& Khangulyan, D. 2012, Nature, 482, 507

\bibitem[{Balbo {et~al.}(2011)Balbo, Walter, Ferrigno, \& Bordas}]{Balbo2011}
Balbo, M., Walter, R., Ferrigno, C., \& Bordas, P. 2011, Astronomy {\&}
  Astrophysics, 527, L4

\bibitem[{Bednarek \& Idec(2011)}]{Bednarek2011}
Bednarek, W., \& Idec, W. 2011, Monthly Notices of the Royal Astronomical
  Society, 414, 2229

\bibitem[{Bhattacharjee {et~al.}(2009)Bhattacharjee, Huang, Yang, \&
  Rogers}]{Bhattacharjee2009}
Bhattacharjee, A., Huang, Y.-M., Yang, H., \& Rogers, B. 2009, Physics of
  Plasmas, 16, 112102

\bibitem[{Bietenholz {et~al.}(2014)Bietenholz, Yuan, Buehler, Lobanov, \&
  Blandford}]{Bietenholz2014}
Bietenholz, M.~F., Yuan, Y., Buehler, R., Lobanov, A.~P., \& Blandford, R.
  2014, Monthly Notices of the Royal Astronomical Society, 446, 205

\bibitem[{Biskamp(1986)}]{Biskamp1986}
Biskamp, D. 1986, Physics of Fluids, 29, 1520

\bibitem[{Blandford {et~al.}(2015{\natexlab{a}})Blandford, East, Nalewajko,
  Yuan, \& Zrake}]{Blandford2015a}
Blandford, R., East, W., Nalewajko, K., Yuan, Y., \& Zrake, J.
  2015{\natexlab{a}}, arXiv:1511.07515

\bibitem[{Blandford {et~al.}(2015{\natexlab{b}})Blandford, Yuan, \&
  Zrake}]{Blandford2015}
Blandford, R., Yuan, Y., \& Zrake, J. 2015{\natexlab{b}}, American Astronomical
  Society

\bibitem[{Bogovalov(1999)}]{Bogovalov1999}
Bogovalov, S.~V. 1999, Astronomy and Astrophysics

\bibitem[{Brandenburg {et~al.}(2015)Brandenburg, Kahniashvili, \&
  Tevzadze}]{Brandenburg2015}
Brandenburg, A., Kahniashvili, T., \& Tevzadze, A.~G. 2015, Physical Review
  Letters, 114, 075001

\bibitem[{Bucciantini {et~al.}(2011)Bucciantini, Arons, \&
  Amato}]{Bucciantini2011}
Bucciantini, N., Arons, J., \& Amato, E. 2011, Monthly Notices of the Royal
  Astronomical Society, 410, 381

\bibitem[{Buehler {et~al.}(2012)Buehler, Scargle, Blandford, Baldini, Baring,
  Belfiore, Charles, Chiang, D'Ammando, Dermer, Funk, Grove, Harding, Hays,
  Kerr, Massaro, Mazziotta, Romani, {Saz Parkinson}, Tennant, \&
  Weisskopf}]{Buehler2012}
Buehler, R., Scargle, J.~D., Blandford, R.~D., {et~al.} 2012, The Astrophysical
  Journal, 749, 26

\bibitem[{Bykov {et~al.}(2012)Bykov, Pavlov, Artemyev, \& Uvarov}]{Bykov2012}
Bykov, A.~M., Pavlov, G.~G., Artemyev, A.~V., \& Uvarov, Y.~A. 2012, Monthly
  Notices of the Royal Astronomical Society: Letters, 421, L67

\bibitem[{Camus {et~al.}(2009)Camus, Komissarov, Bucciantini, \&
  Hughes}]{Camus2009}
Camus, N.~F., Komissarov, S.~S., Bucciantini, N., \& Hughes, P.~A. 2009,
  Monthly Notices of the Royal Astronomical Society, 400, 1241

\bibitem[{Cerutti {et~al.}(2015)Cerutti, Philippov, Parfrey, \&
  Spitkovsky}]{Cerutti2015}
Cerutti, B., Philippov, A., Parfrey, K., \& Spitkovsky, A. 2015, Monthly
  Notices of the Royal Astronomical Society, 448, 606

\bibitem[{Cerutti {et~al.}(2014{\natexlab{a}})Cerutti, Werner, Uzdensky, \&
  Begelman}]{Cerutti2014}
Cerutti, B., Werner, G.~R., Uzdensky, D.~A., \& Begelman, M.~C.
  2014{\natexlab{a}}, Physics of Plasmas, 21, 056501

\bibitem[{Cerutti {et~al.}(2014{\natexlab{b}})Cerutti, Werner, Uzdensky, \&
  Begelman}]{Cerutti2014a}
---. 2014{\natexlab{b}}, The Astrophysical Journal, 782, 104

\bibitem[{Childress(1970)}]{Childress1970}
Childress, S. 1970, Journal of Mathematical Physics, 11, 3063

\bibitem[{Clausen-Brown \& Lyutikov(2012)}]{Clausen-Brown2012}
Clausen-Brown, E., \& Lyutikov, M. 2012, Monthly Notices of the Royal
  Astronomical Society, 426, 1374

\bibitem[{Coroniti(1990)}]{Coroniti1990}
Coroniti, F.~V. 1990, The Astrophysical Journal, 349, 538

\bibitem[{Dombre {et~al.}(1986)Dombre, Frisch, Greene, H{\'{e}}non, Mehr, \&
  Soward}]{Dombre1986}
Dombre, T., Frisch, U., Greene, J.~M., {et~al.} 1986, Journal of Fluid
  Mechanics, 167, 353

\bibitem[{East {et~al.}(2015)East, Zrake, Yuan, \& Blandford}]{East2015}
East, W.~E., Zrake, J., Yuan, Y., \& Blandford, R.~D. 2015, Physical Review
  Letters, 115, 095002

\bibitem[{Emmering \& Chevalier(1987)}]{Emmering1987}
Emmering, R.~T., \& Chevalier, R.~A. 1987, The Astrophysical Journal, 321, 334

\bibitem[{Guo {et~al.}(2015)Guo, Liu, Daughton, \& Li}]{Guo2015}
Guo, F., Liu, Y.-H., Daughton, W., \& Li, H. 2015, The Astrophysical Journal,
  806, 167

\bibitem[{Harris(1962)}]{Harris1962}
Harris, E.~G. 1962, Il Nuovo Cimento, 23, 115

\bibitem[{Hoshino(2012)}]{Hoshino2012}
Hoshino, M. 2012, Physical review letters, 108, 135003

\bibitem[{Huang \& Bhattacharjee(2010)}]{Huang2010}
Huang, Y.-M., \& Bhattacharjee, A. 2010, Physics of Plasmas, 17, 062104

\bibitem[{Inoue(2012)}]{Inoue2012}
Inoue, T. 2012, The Astrophysical Journal, 760, 43

\bibitem[{Kelner {et~al.}(2013)Kelner, Aharonian, \& Khangulyan}]{Kelner2013}
Kelner, S.~R., Aharonian, F.~A., \& Khangulyan, D. 2013, The Astrophysical
  Journal, 774, 61

\bibitem[{Kennel \& Coroniti(1984)}]{Kennel1984a}
Kennel, C.~F., \& Coroniti, F.~V. 1984, The Astrophysical Journal, 283, 710

\bibitem[{Kirk \& Skjaraasen(2003)}]{Kirk2003}
Kirk, J.~G., \& Skjaraasen, O. 2003, The Astrophysical Journal, 591, 366

\bibitem[{Komissarov(2012)}]{Komissarov2012}
Komissarov, S.~S. 2012, Monthly Notices of the Royal Astronomical Society, 428,
  2459

\bibitem[{Kouzu {et~al.}(2013)Kouzu, {S. Tashiro}, Terada, Yamada, Bamba,
  Enoto, Mori, Fukazawa, \& Makishima}]{Kouzu2013}
Kouzu, T., {S. Tashiro}, M., Terada, Y., {et~al.} 2013, Publications of the
  Astronomical Society of Japan, 65, 74

\bibitem[{Lyubarsky \& Kirk(2001)}]{Lyubarsky2001}
Lyubarsky, Y., \& Kirk, J.~G. 2001, The Astrophysical Journal, 547, 437

\bibitem[{Lyubarsky(2012)}]{Lyubarsky2012}
Lyubarsky, Y.~E. 2012, Monthly Notices of the Royal Astronomical Society, 427,
  1497

\bibitem[{Madsen {et~al.}(2015)Madsen, Reynolds, Harrison, An, Boggs,
  Christensen, Craig, Fryer, Grefenstette, Hailey, Markwardt, Nynka, Stern,
  Zoglauer, \& Zhang}]{Madsen2015}
Madsen, K.~K., Reynolds, S., Harrison, F., {et~al.} 2015, The Astrophysical
  Journal, 801, 66

\bibitem[{McKinney(2006)}]{McKinney2006a}
McKinney, J.~C. 2006, Monthly Notices of the Royal Astronomical Society, 367,
  1797

\bibitem[{Meyer {et~al.}(2010)Meyer, Horns, \& Zechlin}]{Meyer2010}
Meyer, M., Horns, D., \& Zechlin, H.-S. 2010, Astronomy {\&} Astrophysics, 523,
  A2

\bibitem[{Michel(1973)}]{Michel1973}
Michel, F.~C. 1973, The Astrophysical Journal, 180, 207

\bibitem[{Michel(1994)}]{Michel1994}
---. 1994, The Astrophysical Journal, 431, 397

\bibitem[{Olesen(2015)}]{Olesen2015}
Olesen, P. 2015, arXiv:1509.08962

\bibitem[{Parker(1958)}]{Parker1958}
Parker, E.~N. 1958, The Astrophysical Journal, 128, 664

\bibitem[{Porth {et~al.}(2013)Porth, Komissarov, \& Keppens}]{Porth2013}
Porth, O., Komissarov, S.~S., \& Keppens, R. 2013, Monthly Notices of the Royal
  Astronomical Society, 438, 278

\bibitem[{Radice \& Rezzolla(2013)}]{Radice2013}
Radice, D., \& Rezzolla, L. 2013, The Astrophysical Journal, 766, L10

\bibitem[{Rees \& Gunn(1974)}]{Rees1974}
Rees, M.~J., \& Gunn, J.~E. 1974, Monthly Notices of the Royal Astronomical
  Society, 167, 1

\bibitem[{Rudy {et~al.}(2015)Rudy, Horns, DeLuca, Kolodziejczak, Tennant, Yuan,
  Buehler, Arons, Blandford, Caraveo, Costa, Funk, Hays, Lobanov, Max, Mayer,
  Mignani, O’Dell, Romani, Tavani, \& Weisskopf}]{Rudy2015}
Rudy, A., Horns, D., DeLuca, A., {et~al.} 2015, The Astrophysical Journal, 811,
  24

\bibitem[{Rybicki \& Lightman(1979)}]{Rybicki1979}
Rybicki, G.~B., \& Lightman, A.~P. 1979, New York

\bibitem[{Sironi \& Spitkovsky(2011)}]{Sironi2011}
Sironi, L., \& Spitkovsky, A. 2011, The Astrophysical Journal, 741, 39

\bibitem[{Sironi \& Spitkovsky(2014)}]{Sironi2014}
---. 2014, The Astrophysical Journal, 783, L21

\bibitem[{Striani {et~al.}(2013)Striani, Tavani, Vittorini, Donnarumma,
  Giuliani, Pucella, Argan, Bulgarelli, Colafrancesco, Cardillo, Costa, {Del
  Monte}, Ferrari, Mereghetti, Pacciani, Pellizzoni, Piano, Pittori, Rapisarda,
  Sabatini, Soffitta, Trifoglio, Trois, Vercellone, \&
  Verrecchia}]{Striani2013}
Striani, E., Tavani, M., Vittorini, V., {et~al.} 2013, The Astrophysical
  Journal, 765, 52

\bibitem[{Sturrock \& Aschwanden(2012)}]{Sturrock2012}
Sturrock, P., \& Aschwanden, M.~J. 2012, The Astrophysical Journal, 751, L32

\bibitem[{Tanaka \& Takahara(2010)}]{Tanaka2010}
Tanaka, S.~J., \& Takahara, F. 2010, The Astrophysical Journal, 715, 1248

\bibitem[{Tavani {et~al.}(2011)Tavani, Bulgarelli, Vittorini, Pellizzoni,
  Striani, Caraveo, Weisskopf, Tennant, Pucella, Trois, Costa, Evangelista,
  Pittori, Verrecchia, {Del Monte}, Campana, Pilia, {De Luca}, Donnarumma,
  Horns, Ferrigno, Heinke, Trifoglio, Gianotti, Vercellone, Argan, Barbiellini,
  Cattaneo, Chen, Contessi, D'Ammando, DePris, {Di Cocco}, {Di Persio}, Feroci,
  Ferrari, Galli, Giuliani, Giusti, Labanti, Lapshov, Lazzarotto, Lipari,
  Longo, Fuschino, Marisaldi, Mereghetti, Morelli, Moretti, Morselli, Pacciani,
  Perotti, Piano, Picozza, Prest, Rapisarda, Rappoldi, Rubini, Sabatini,
  Soffitta, Vallazza, Zambra, Zanello, Lucarelli, Santolamazza, Giommi,
  Salotti, \& Bignami}]{Tavani2011}
Tavani, M., Bulgarelli, A., Vittorini, V., {et~al.} 2011, Science (New York,
  N.Y.), 331, 736

\bibitem[{Taylor(1974)}]{Taylor1974}
Taylor, J.~B. 1974, Physical Review Letters, 33, 1139

\bibitem[{Teraki \& Takahara(2013)}]{Teraki2013}
Teraki, Y., \& Takahara, F. 2013, The Astrophysical Journal, 763, 131

\bibitem[{Uzdensky {et~al.}(2011)Uzdensky, Cerutti, \& Begelman}]{Uzdensky2011}
Uzdensky, D.~A., Cerutti, B., \& Begelman, M.~C. 2011, The Astrophysical
  Journal, 737, L40

\bibitem[{Uzdensky {et~al.}(2010)Uzdensky, Loureiro, \&
  Schekochihin}]{Uzdensky2010}
Uzdensky, D.~A., Loureiro, N.~F., \& Schekochihin, A.~A. 2010, Physical review
  letters, 105, 235002

\bibitem[{Uzdensky \& Spitkovsky(2014)}]{Uzdensky2014}
Uzdensky, D.~A., \& Spitkovsky, A. 2014, The Astrophysical Journal, 780, 3

\bibitem[{Zelenyi \& Krasnosel'skikh(1979)}]{Zelenyi1979}
Zelenyi, L.~M., \& Krasnosel'skikh, V.~V. 1979, Soviet Astronomy, 23

\bibitem[{Zrake(2014)}]{Zrake2014}
Zrake, J. 2014, The Astrophysical Journal, 794, L26

\bibitem[{Zrake \& East(2015)}]{Zrake2015}
Zrake, J., \& East, W.~E. 2015, arXiv:1509.00461

\bibitem[{Zrake \& MacFadyen(2012)}]{Zrake2012}
Zrake, J., \& MacFadyen, A. 2012, The Astrophysical Journal Letters

\bibitem[{Zrake \& MacFadyen(2011)}]{Zrake2011}
Zrake, J., \& MacFadyen, A.~I. 2011, The Astrophysical Journal, 744, 32

\end{thebibliography}

\end{document}